\documentclass[12pt, a4paper]{article} 
\usepackage[includeheadfoot,
            marginratio={1:1,2:3}, 
            width=452pt, 
            height=720pt,]{geometry}

\usepackage{amsmath}
\usepackage{amsfonts}
\usepackage{amssymb}
\usepackage{bbm,bm}
\usepackage{mathtools}
\usepackage{slashed}
\usepackage{mathalfa}
\usepackage{graphicx,caption,subfigure}
\usepackage{tikz} 
\usepackage{tikz-cd}
\usetikzlibrary{decorations.pathreplacing,calc,tikzmark}
\usepackage{booktabs}
\usepackage{adjustbox}
\usepackage[utf8]{inputenc}
 \usepackage[splitrule]{footmisc}

\usepackage{placeins}
\usepackage{empheq}
\usepackage{paralist}
\usepackage{cite}
\usepackage[normalem]{ulem}
\usepackage{soul}

\usepackage[colorlinks=false,urlbordercolor=red
]{hyperref}
\usepackage{xcolor}
\usepackage{tensor}

\newcommand{\nc}{\newcommand}
\nc{\lb}{\llbracket}
\nc{\rb}{\rrbracket}
\nc{\gl}{\llbracket}
\nc{\gr}{\rrbracket}
\nc{\del}{\partial}
\nc{\tri}{\hspace{-3.5pt}\vartriangle\hspace{-3.5pt}}
\nc{\blacktri}{\blacktriangle}

\nc{\eq}[1]{\begin{equation}
                     \begin{split} #1 \end{split}
                     \end{equation}}
\nc{\ov}{\overline}

\nc{\fa}{\hat}
\nc{\fb}{\MakeUppercase}
\nc{\fc}{\tilde }
\nc{\Lie}{{\cal L}} 
\nc{\lambdabar}{{\mkern0.75mu\mathchar '26\mkern -9.75mu\lambda}}

\allowdisplaybreaks[2]
\numberwithin{equation}{section}

\DeclareUnicodeCharacter{2212}{-}
\begin{document}

\vspace*{-1.5cm}
\begin{flushright}
  {\small
  MPP-2023-45\\
  }
\end{flushright}

\vspace{1.5cm}
\begin{center}
  {\Large \bf Quantum gravity constraints on scale separation\\[.5cm] and de Sitter in five dimensions} 
\vspace{0.35cm}

\end{center}

\vspace{0.35cm}
\begin{center}
{\large 
Niccol\`o Cribiori$^a$ and Carmine Montella$^b$
}
\end{center}

\vspace{0.1cm}
\begin{center} 
\emph{
$^a$Max-Planck-Institut f\"ur Physik (Werner-Heisenberg-Institut), \\[.1cm] 
   F\"ohringer Ring 6,  80805 M\"unchen, Germany, 
   \\[0.1cm] 
 \vspace{0.3cm}
$^b$Dipartimento di Fisica e Astronomia, Universit\`a di Bologna, \\
via Irnerio 46, 40126 Bologna, Italy, \\[.1cm] 
    } 
\end{center} 

\vspace{0.5cm}

\begin{abstract}

We give evidence that supersymmetric anti-de Sitter vacua of five-dimensional supergravity cannot be scale separated as a consequence of quantum gravity constraints, such as the weak gravity conjecture or the species scale. We show this in a model-independent way for the minimal and the maximal theory and we believe that the argument can be extended to any amount of preserved supercharges in the between. If combined with previous works stating that non-supersymmetric anti-de Sitter vacua must be unstable, our results suggest that five-dimensional effective field theories in anti-de Sitter belong to the swampland. At the cost of introducing an additional assumption on the gravitino mass, we can extend our analysis to de Sitter vacua as well. However, the few known stable de Sitter vacua of minimal five-dimensional supergravity do not satisfy such an assumption and thus evade our constraints. 
This suggest that they are on a somehow different footing than their four-dimensional counterparts and therefore deserve further investigation.

\end{abstract}

\thispagestyle{empty}
\clearpage

\setcounter{tocdepth}{2}

\tableofcontents

%\newpage

\section{Introduction}

At present, the Swampland program \cite{Vafa:2005ui,Palti:2019pca,vanBeest:2021lhn,Agmon:2022thq} collects a series of consistency constraints that low energy effective field theories have to satisfy in order to be compatible with quantum gravity. The scope is ambitious, for one would like to use these criteria to motivate and explain phenomenological properties of the universe. Two examples are the fact that no extra dimensions have been detected so far and that a positive value of dark energy density is measured today. These properties are the focus of the present work.

One of the reasons why no extra dimensions are detected can be that they are too small to be probed with the current technology. This is a common assumption in the context of string compactifications, where one starts from critical string theory and then needs to extract a lower dimensional effective theory.\footnote{Alternative scenarios are possible as well, such as those with (undetected) large extra dimensions; see e.g.~\cite{Montero:2022prj,Danielsson:2022lsl} for recent proposals. Nevertheless, here we will focus on the case in which the extra dimensions are small and compact.} 
Whether or not this procedure is capable of producing an effective description which is genuinely lower dimensional is not completely settled, especially on a curved background. Indeed, in the best understood case of supersymmetric anti-de Sitter vacua, the majority of the models seem not to allow for a separation of scales between the sizes of the compact and non-compact dimensions, see for example \cite{Lust:2004ig,Tsimpis:2012tu,Lust:2020npd,DeLuca:2021mcj,Collins:2022nux,Andriot:2022yyj}. This has been encoded into Swampland conjectures \cite{Lust:2019zwm}, see also \cite{Blumenhagen:2019vgj,Buratti:2020kda,Apers:2022zjx}. However, counterexamples are present as well, such as the four-dimensional class of models in massive IIA compactifications \cite{DeWolfe:2005uu} (based on earlier works \cite{Behrndt:2004km,Derendinger:2004jn}), which recently passed non-trivial consistency checks \cite{Junghans:2020acz,Marchesano:2020qvg,Shiu:2022oti}, a (formal) double T-dual version in massless IIA and M-theory \cite{Cribiori:2021djm}, and the three-dimensional models of \cite{Farakos:2020phe,Emelin:2022cac,VanHemelryck:2022ynr}. 

Given that the situation is unclear, one can try to make progress by remaining agnostic about any swampland conjecture directly forbidding scale separation, and instead demonstrating that scale separation is not possible as a consequence of other conjectures or principles. For four-dimensional anti-de Sitter vacua with at least eight preserved supercharges, this logic has been pursued in \cite{Cribiori:2022trc} by using the weak gravity conjecture \cite{Arkani-Hamed:2006emk}, and it has been extended further in \cite{Montero:2022ghl}, by enforcing the absence of global symmetries.
In this work, we generalize the analysis of \cite{Cribiori:2022trc} to five dimensions, relaxing some of the hypothesis therein, and eventually we make contact with \cite{Montero:2022ghl}. 
Furthermore, we provide an argument against scale separation which is based on holography and on the concept of species scale \cite{Dvali:2007wp,Dvali:2007hz,Dvali:2010vm}.

The step from four to five dimensions is more than an exercise, for in five dimensions the minimal spinor contains already eight supercharges and thus, contrary to four dimensions, one can provide general statements even for vacua with minimal supersymmetry.\footnote{Swampland conjectures in five dimensions have been studied recently also in \cite{Heidenreich:2020ptx,Katz:2020ewz,Cota:2022maf,Gendler:2022ztv}.}
Indeed, we show in a model-independent way that scale separation is not possible in minimally and maximally supersymmetric anti-de Sitter vacua in five dimensions, and we see no obstruction in extending the analysis to vacua with intermediate supersymmetry. When combining our findings with \cite{Ooguri:2016pdq}, arguing that non-supersymmetric anti-de Sitter vacua are unstable, one is led to conclude that anti-de Sitter vacua in five dimensions are never consistent lower dimensional effective theories and rather belong to the swampland, regardless of the number of preserved supercharges.

The second property of effective field theories that is motivated by observations and that we investigate in the present work is the presence of de Sitter vacua.\footnote{A de Sitter phase is not the only possible explanation for the positive value of dark energy density measured today. For example, one could also consider quintessence models. Nevertheless, the focus of the present work is on de Sitter critical points of the scalar potential.} As for scale separation, whether or not de Sitter vacua can be realized in string theory is an open problem. On the one hand, there are celebrated scenarios such as \cite{Kachru:2003aw,Balasubramanian:2005zx}, on the other hand criticisms have been raised, see e.g.~\cite{Gao:2020xqh,Junghans:2022exo,Lust:2022lfc,Junghans:2022kxg,Blumenhagen:2022dbo} for recent works. In the context of the Swampland program, conjectures have been formulated which constrain or even forbid de Sitter critical points, see for example \cite{Obied:2018sgi,Andriot:2018wzk,Garg:2018reu,Ooguri:2018wrx,Andriot:2018mav,Rudelius:2019cfh,Bedroya:2019snp}. However, perhaps with the exception of \cite{Bedroya:2019snp}, the underlying motivation for these conjectures seems to be mainly the lack of explicit examples, but it is not yet clear if de Sitter is incompatible with quantum gravity at a more fundamental level.

To make progress, one can again try to constrain de Sitter vacua by means of other, more established, swampland ideas. 
This logic has been pursued in \cite{Cribiori:2020use,DallAgata:2021nnr} (see also \cite{Cribiori:2020wch}), where the weak gravity conjecture has been employed to argue that four-dimensional de Sitter critical points of extended supergravity with parametrically small or vanishing gravitino mass are in the swampland.\footnote{In the present work, we always refer to the lagrangian gravitino mass.} 
In particular, this holds true for all known stable de Sitter solutions of $\mathcal{N}=2$ supergravity and also for large classes of unstable ones, since they feature a vanishing gravitino mass.  
In this work, we extend the analysis of \cite{Cribiori:2020use,DallAgata:2021nnr} to five dimensions, where we can provide non-trivial statements for the minimal theory.

Essentially, the core of our analysis is that the cosmological constant of supersymmetric anti-de Sitter vacua, or of de Sitter vacua with parametrically small gravitino mass, is such that (in Planck units and suppressing constant factors)
\begin{equation}
\label{Vgtrg}
|V| \geq g_{3/2}^2,
\end{equation}
where $g_{3/2}$ is the gravitino gauge coupling associated to the gauging of the R-symmetry. This happens for all gauged supergravities with at least eight supercharges, regardless of the matter content, and the relation is actually saturated for anti-de Sitter vacua with eight preserved supercharges.  
We will argue that models satisfying \eqref{Vgtrg} cannot be genuinely five-dimensional. They are instead higher dimensional for they do not admit a separation of scales between compact and non-compact dimensions. 
For supersymmetric anti-de Sitter vacua, our arguments are general and model-independent. Hence, they have implications for holography that would be interesting to investigate in the future.
For de Sitter vacua, additional assumptions are needed and thus our conclusions are less general. In particular, in contrast to the four-dimensional analysis of \cite{Cribiori:2020use,DallAgata:2021nnr}, in five dimension we will not be able to constrain the few known stable de Sitter vacua \cite{Cosemans:2005sj, Ogetbil:2008tk,Ogetbil:2008ojt}, for they typically feature a non-vanishing gravitino mass. The fate of these vacua remains thus as an open problem.

\section{Constraints on minimal supergravity}

We start by considering minimal supergravity in five dimensions. We first review the main ingredients of the theory and then discuss swampland constraints on scale separation in supersymmetric anti-de Sitter vacua and on the effective field theory describing de Sitter vacua. For the first kind of vacua, we also provide an argument against scale separation based on holography and on the notion of species scale. 

Our approach is exclusively bottom-up. This implies that, no matter what the precise details of the ultraviolet completion are, if a given microscopic model admits an effective description with an anti-de Sitter (or de Sitter) vacuum with at least eight preserved supercharges (at the lagrangian level), our results apply.

\subsection{Minimal supergravity in five dimensions}
\label{sec:revminsugra}

The general matter coupled gauged theory was constructed in \cite{Ceresole:2000jd} (see \cite{Gunaydin:1983bi,Gunaydin:1999zx,Gunaydin:2000xk,Gunaydin:2000ph} for earlier works) and subsequently revisited in \cite{Bergshoeff:2004kh}. 
A recent textbook on the subject is \cite{Lauria:2020rhc}. 
Here, we mainly follow the conventions of \cite{Bergshoeff:2004kh}. 

The ingredients are the gravity multiplet and the so-called matter multiplets, namely $n_V$ vector, $n_T$ tensor and $n_H$ hyper multiplets.\footnote{In the ungauged theory, tensor multiplets can always be dualized to vector multiplets and viceversa. In the gauged theory, this is in general not true. Indeed, as it was originally noticed in maximal supergravity in seven dimensions \cite{Pernici:1984xx}, tensor multiplets have to be introduced whenever vector fields not actively participating to the gauging are nevertheless charged. More precisely, to preserve supersymmetry, either one performs a Higgs mechanism to give mass to the vectors or, in odd dimensions, one can dualize such vectors to tensors. For minimal supergravity in five dimensions, this is explained in \cite{Gunaydin:1999zx}.} 
The gravity multiplet contains the graviton, an SU(2)$_R$-doublet of symplectic-Majorana gravitini $\psi^i_\mu$, with $i=1,2$, and the graviphoton. 
Each vector multiplet contains one vector, an SU(2)$_R$-doublet of spin-1/2 fermions $\lambda^i$ and one real scalar. 
Each tensor multiplet contains a rank-two antisymmetric tensor, an SU(2)$_R$-doublet of spin-1/2 fermions and one real scalar.
Each hyper multiplet contains a doublet of spin-1/2 fermions $\lambda^A$, with $A=1,2$ index of USp$(2)$, and four real scalars $q^u$, with $u=1,\dots,4$. 
It is convenient to collect all of the real scalars from vector and tensor multiplets into a single object, $\phi^{\tilde x}$, with $\tilde x=1,\dots, n_V+n_T$. 
Similarly, one introduces a collective index for the vector and tensor fields, $\tilde I=(I,M)$, with $I=0,\dots,n_V$ and $M=1,\dots,n_T$. 
Notice that the R-symmetry group of the theory is SU$(2)_R$ and not SU$(2)_R \times {\rm U}(1)_R$ as in the four-dimensional case. 
The absence of the U$(1)_R$ factor is related to the fact that scalar fields in vector and tensor multiplets are real. In general, this has important consequence on the microscopic model \cite{Witten:1996qb}, and it makes some aspects of the analysis more transparent.

The full scalar manifold is given by the product
\begin{equation}
\label{Mscalar}
\mathcal{M} = \mathcal{S}(n_V+n_T) \otimes \mathcal{Q}(n_H),
\end{equation}
where $\mathcal{S}$ is an $(n_V+n_T)$-dimensional very special manifold while $\mathcal{Q}$ is a $4 n_H$-dimensional quaternionic K\"ahler manifold. 
The quaternionic geometry is analogous to the four-dimensional case and we thus refer to \cite{Andrianopoli:1996cm} for more details.

The very special manifold $\mathcal{S}$ can be conveniently described as an hypersurface in a larger ambient space. Locally, it is parametrized by coordinates $h^{\tilde I}(\phi^{\tilde x})$ subject to the constraint
\begin{equation}
\label{F=1constr}
\mathcal{F}(h)=  C_{\tilde I\tilde J \tilde K} h^{\tilde I}h^{\tilde J}h^{\tilde K} = 1,
\end{equation}
where $C_{\tilde I \tilde J \tilde K}$ is constant and completely symmetric. 
The metric on the ambient space is given by \cite{Gunaydin:1983bi}
\begin{equation}
\begin{aligned}
\label{aijdef}
a_{\tilde I \tilde J} = - \frac13\partial_{\tilde I} \partial_{\tilde J} \log \mathcal{F}(h)\,|_{\mathcal{F}=1}=- 2C_{\tilde I \tilde J \tilde K} h^{\tilde K}+ 3h_{\tilde I}h_{\tilde J},
\end{aligned}
\end{equation}
furthermore one has $h_{\tilde I} = C_{\tilde I \tilde J \tilde K}h^{\tilde J}h^{\tilde K} =a_{\tilde I \tilde J}h^{\tilde J}$ and $h^{\tilde I} h_{\tilde I}=1$. It is useful to introduce the vielbeins
\begin{equation}
\label{hxihix}
h_{\tilde x}^{\tilde I} = -\sqrt{\frac 32} \,\partial_{\tilde x}h^{\tilde I}(\phi), \qquad h_{\tilde I\,\tilde x} = \sqrt{\frac 32} \,\partial_{\tilde x}h_{\tilde I}(\phi),
\end{equation}
obeying the orthogonality relation
\begin{equation}
\label{hix}
h^{\tilde I}_{\tilde x} h_{\tilde I} = h_{\tilde I\,\tilde x}h^{\tilde I}=0.
\end{equation}
One can use these vielbeins to write the metric of the very special manifold $\mathcal{S}$ as
\begin{equation}
\label{gxy}
g_{\tilde x \tilde y} = h^{\tilde I}_{\tilde x} h^{\tilde J}_{\tilde y} a_{\tilde I\tilde J},
\end{equation}
and one can check that
\begin{equation}
\label{aij=hh}
 h_{\tilde I} h_{\tilde J} + g_{\tilde x \tilde y} h_{\tilde I}^{\tilde x} h_{\tilde J}^{\tilde y} = a_{\tilde I\tilde J} .
\end{equation}
The curved indices $\tilde{x}$, $\tilde{y}$, $\dots$ are lowered and raised with the metric $g_{\tilde x \tilde y}$, while $\tilde I$, $\tilde J$, $\dots$ with $a_{\tilde I \tilde J}$. Useful relations in very special geometry can be found e.g.~in \cite{Ceresole:2000jd,Bergshoeff:2004kh}.

The metric $a_{\tilde I \tilde J}$ enters the action directly via the kinetic term of the vector and tensor fields, which reads
\begin{equation}
\label{kinvec}
e^{-1}\mathcal{L}_{\text{kin}}^{\text vec} =-\frac14 a_{\tilde I \tilde J}  \mathcal{H}^{\tilde I}_{\mu\nu}  \mathcal{H}^{\tilde J\, \mu\nu},
\end{equation}
where $\mathcal{H}^{\tilde I}_{\mu\nu}$ collects both vector field strengths and tensors, namely $\mathcal{H}^{\tilde I}_{\mu\nu} = (F^I_{\mu\nu}, B_{\mu\nu}^M)$, with $F^{I}_{\mu\nu}=2\partial_{[\mu}A_{\nu]}^I+{f_{JK}}^IA^J_\mu A^K_\nu$ and ${f_{JK}}^I$ the structure constants of the gauge algebra. In order to avoid ghosts, the matrix $a_{\tilde I \tilde J}$ has to be positive semi-definite. Similarly, the metric $g_{\tilde x \tilde y}$ enters the kinetic term of the very special scalars,
\begin{equation}
e^{-1}\mathcal{L}_{\text{kin}}^{\mathcal{S}} = -\frac12 g_{\tilde x \tilde y}D_\mu \phi^{\tilde x} D^\mu \phi^{\tilde y},
\end{equation}
and it has to be positive semi-definite as well.

The ungauged theory stems from M-theory compactified on a Calabi-Yau threefold \cite{Cadavid:1995bk}. In this case, $C_{IJK}$ with $I,J,K=1,\dots,h^{1,1}(CY_3)$ is the triple intersection number and the constraint \eqref{F=1constr} expresses the fact that the Calabi-Yau volume is described by the universal hyper multiplet and not by a vector multiplet. The vector fields arise from the Kaluza-Klein reduction of the M-theory three-form on a basis of two-forms of the Calabi-Yau. The scalar fields in the vector multiplets are related to fluctuations of the K\"ahler two-form, while those in the hyper multiplets to fluctuations of the holomorphic three-form.

From the five-dimensional perspective, the gauged theory can be constructed as a deformation of the ungauged one in which new terms appear both in the action and in the supersymmetry transformations of the fermions (and of the tensors). 
Remarkably, the Noether procedure stops at $\mathcal{O}(g^2)$, $g$ being a proxy for the gauge coupling, and the term in the action containing $g^2$ is the scalar potential.

In practice, the gauging is performed by introducing two sets of Killing vectors, $K^{\tilde x}_I(\phi)$ and $K^u_I(q)$, which are functions on $\mathcal{S}$ and $\mathcal{Q}$ respectively and generate the isometries
\begin{align}
\label{Killingx}
\delta \phi^{\tilde x} &= \theta^I K_I^{\tilde x}(\phi),\\
\delta q^{u} &= \theta^I K_I^{u}(\phi),
\end{align}
with $\theta^I$ infinitesimal real parameters.
The very special Killing vectors $K^{\tilde x}_I (\phi)$ are given by the expression
\begin{equation}
\label{Killingx2}
K^{\tilde x}_I(\phi) = -\sqrt \frac32 {t_{I\tilde J}}^{\tilde K} h^{\tilde J} h^{\tilde x}_{\tilde K} = -\sqrt \frac32 {t_{I\tilde J}}^{\tilde K} h^{\tilde J\, \tilde x} h_{\tilde K},
\end{equation}
where ${t_{I\tilde J}}^{\tilde K}$ are generalizations of the structure constants ${f_{IJ}}^K$ of the gauge algebra and have to obey certain constraints due to supersymmetry \cite{Bergshoeff:2002qk,Bergshoeff:2004kh}.
The quaternionic Killing vectors are the (covariant) derivatives of an SU$(2)_R$ triplet of quaternionic prepotentials $P_I^{\rm r}(q)$, which typically appear contracted with the Pauli matrices $\sigma^{\rm r}$, with ${\rm r}=1,2,3$,
 \begin{equation}
{P_{I\, i}}^j(q) = i P_I^{\rm r} {(\sigma^{\rm r})_i}^j .
\end{equation}

Due to the gauging, the supersymmetry transformations of the fermions receive a modification of the type
\begin{align}
\label{deltagaugino}
\delta\lambda^{\tilde x\, i} &=\dots -g_R P^{\tilde x\,ij} \epsilon_j+g W^{\tilde x}\epsilon^i ,\\
\label{deltahyper}
\delta \zeta^A &= \dots + g \mathcal{N}_i^A \epsilon^i,\\
\label{deltagrav}
\delta \psi_\mu^i &= D_\mu \epsilon^i - \frac{1}{\sqrt 6}ig_R P^{ij}\gamma_\mu \epsilon_j+\dots,
\end{align}
where dots stand for terms that are present also in the ungauged theory but that are not relevant for us, for they vanish on maximally symmetric vacua. Here, following \cite{Ceresole:2000jd}, we are making a distinction between the gauge coupling $g$, associated to local isometries of the scalar manifold, and the R-symmetry gauge coupling $g_R$.
Explicitly, the gauging terms in \eqref{deltagaugino}-\eqref{deltagrav} contain the quantities
\begin{align}
P_{ij} &= h^I P_I^{ij},\\
P_{ij}^{\tilde x} &=h^{\tilde x\, I}P_{I\, ij},\\
W^{\tilde x} &=  \frac{\sqrt 6}{4}h^I K_I^{\tilde x},\\
\mathcal{N}^{iA} &= \frac{\sqrt 6}{4}h^I K_I^u f^{iA}_u ,
\end{align}
where $f^{iA}_u$ is a vielbein on the quaternionic manifold.
These determine completely the scalar potential, which reads
\begin{equation}
\label{VN=2gen}
V =g_R^2 \left(P^{\tilde x}_{ij}P^{ij}_{\tilde x} - 2P_{ij}P^{ij}\right) + 2g^2 W^{\tilde x} W_{\tilde x}+2g^2\mathcal{N}_{iA}\mathcal{N}^{iA}.
\end{equation}
Notice that the last two terms in \eqref{deltagaugino} are independent, for they are associated to fermionic parameters of opposite chirality. Indeed, one has the orthogonality condition \cite{Ceresole:2000jd}
\begin{equation}
\label{wp=0}
W_{\tilde x }P^{\tilde x}_ {ij} =0,
\end{equation}
which can be proven with the help of ${t_{I\tilde J}}^{\tilde K} h^{\tilde J}h_{\tilde K}=0$ and ${t_{JM}}^I=0$.
Finally, we will need to look at the gravitino covariant derivative, which reads
\begin{equation}
\label{gravcovder}
\mathcal{D}_\mu\psi_\nu^i = \left(\partial_\mu+\frac14 \omega_\mu^{ab}\gamma_{ab}\right)\psi_\nu^i - \partial_\mu q^u \omega_u^{ij}\psi_{\nu\, j} - g_R A^I_\mu P_I^{ij}\psi_{\nu\, j}.
\end{equation}
From this formula, one sees that $P_I^{ij}$ has the role of a field-dependent gravitino charge.

\subsection{No scale separation in anti-de Sitter}
\label{sec_noscalesepN=2}

We are going to argue that the absence of scale separation in minimally supersymmetric anti-de Sitter vacua in five-dimensions follows as a consequences of some of the most established swampland conjectures, such as the absence of global symmetries in quantum gravity or the weak gravity conjecture. In some sense, this corresponds to recover the (strong) anti-de Sitter distance conjecture \cite{Lust:2019zwm} from the aforementioned conjectures.

The conditions for a supersymmetric vacuum can be found by setting to zero the supersymmetry variations of the fermions. In our case, this gives
\begin{equation}
\label{susyAdSN=2}
P^{\tilde x \,ij} = W^{\tilde x} =\mathcal{N}_i^A =0
\end{equation}
and the scalar potential reduces to
\begin{equation}
V_{AdS} =-2 g_R^2 P_{ij}P^{ij} = -4 g_R^2 P^{\rm r} P^{\rm r}. 
\end{equation}
Using that $P_{ij} = h^I P_{I\,ij}$, the relation \eqref{aij=hh} and the fact that $P^{\tilde x \,ij} = h^{\tilde x\,I} P_{I\,ij}=0$ on the vacuum, one can rewrite it as
\begin{equation}
\label{VminimalAdS}
V_{AdS}=-2 g_R^2 a^{IJ}P_{I\, ij} P_{J}^{ij} = -4 g_R^2 a^{IJ} P^{\rm r}_I P^{\rm r}_J.
\end{equation}
This equation expresses the fact that the cosmological constant of a supersymmetric anti-de Sitter vacuum is given by the product of the gauge kinetic function $a^{IJ}$ and of the gravitino charge, $P_{I\,ij}$, without any additional parameter entering the relation.

Since a central ingredient in our analysis is the gravitino gauge coupling, associated to the gauging of the R-symmetry, we need to identify it properly. Indeed, $g_R$ appearing in \eqref{VminimalAdS} is just a device introduced to keep track of the gauging deformation. As such, it has no intrinsic physical meaning, besides telling us that the cosmological constant in the supersymmetric vacuum is governed by the R-symmetry gauging. Thus, we proceed by setting $g_R=1$ in what follows and we determine the actual gravitino gauge coupling, which is a function of the scalar fields. 

Actually, we only need to know what the unbroken gauge group on the vacuum and the corresponding gauge coupling are. In general, the answer to this question is model-dependent, for it depends on the choice of the scalar manifold and of the gauge group. However, thanks to supersymmetry, one can provide a fairly universal answer on anti-de Sitter vacua, as done in \cite{Louis:2016qca} for minimal supergravity in five dimensions and extended in \cite{Lust:2017aqj} to a generic number of dimensions and preserved supercharges. There, it is shown that on a maximally supersymmetric anti-de Sitter vacuum the gauge group $G$ is spontaneously broken to a subgroup, which factorizes into two mutually commuting subgroups, $H_R$ and $H_{\rm mat}$, namely\footnote{Notice that we indicate with $H_R$ and $H_{\rm mat}$ the unbroken groups on the vacuum. The same groups are denoted in \cite{Louis:2016qca} with an additional suffix, $H_R^g$ and $H_{\rm mat}^g$, which is here understood. Besides, the former is the R-symmetry of the dual superconformal field theory, while the latter is an additional flavour symmetry. }
\begin{equation}
\label{GtoHRHmat}
G \to H_R \times H_{\rm mat}.
\end{equation}
In particular, the group $H_R$ is gauged by the graviphotons of the theory, while $H_{\rm mat}$ is gauged by vectors in the matter multiplets. In minimal supergravity in five dimensions, which in the language of \cite{Lust:2017aqj} corresponds to $\mathcal{N}=2$, there is only one graviphoton and thus $H_R = {\rm U}(1)_R$. This implies that, on the anti-de Sitter vacua we are interested in the present section, one has always an unbroken abelian factor of the full gauge group.\footnote{In special cases, the gauging of the full SU$(2)_R$ group is also possible, but it is highly constrained and it does not involve the graviphoton. For large classes of models, the resulting scalar potential does not admit critical points \cite{Gunaydin:2000ph}. } 
Let us also notice that the same applies to $\mathcal{N}=2$ anti-de Sitter vacua in four dimensions, where again $H_R =  {\rm U}(1)_R$. Hence, the assumption of \cite{Cribiori:2022trc} about the presence of an unbroken abelian gauge group on the vacuum, which is needed to apply the weak gravity conjecture, is in fact always satisfied with eight supercharges.

Since we have only one graviphoton, the identification of the vector gauging $H_R$ can be performed directly. By looking at the supersymmetry transformation of the gravitini, one sees that the linear combination $h_{\tilde I}\mathcal{H}^{\tilde I}$ of scalar fields and vector field strengths (and tensor) appears. For this reason, the linear combination $h_I A^I$ can be interpreted as the graviphoton of the matter-coupled theory. Similarly, since in the supersymmetry transformations of the gaugini the linear combination $h^{\tilde x}_I \mathcal{H}^{\tilde I}$ appears, the matter vectors are typically identified with $h^{x}_I A^I$. Notice that these linear combinations are orthogonal as a consequence of \eqref{hix}.

Without loss of generality, we thus proceed by considering a linear combination of all vectors (we omit the spacetime index in what follows)
\begin{equation}
\tilde A = \Theta_I A^I, 
\end{equation}
where $\Theta_I$ are arbitrary coefficients, which can be thought of as the vacuum expectation values of $h_I$ appropriately normalized, and we declare that this is gauging an unbroken abelian group on the vacuum. In practice, we can assume it gauges $H_R = {\rm U}(1)_R$. 
Next, we split the whole set of fields $A^I$ into longitudinal and orthogonal directions with respect to $\tilde A$, namely 
\begin{equation}
\begin{aligned}
A_I &= A^\parallel_I + A^\perp_I  = \frac{a^{JK} \Theta_K \Theta_I}{\Theta^2} A_J + A^\perp_I = \frac{\Theta_I}{\Theta^2} \tilde A + A^\perp_I,
\end{aligned}
\end{equation}
where we denoted $\Theta^2 \equiv \Theta_I a^{IJ} \Theta_J$.
Similarly, we can split the following combination entering the gravitino covariant derivative (SU(2)$_R$ indices are understood)
\begin{equation}
\begin{aligned}
A_I P^I &=A_I^\parallel P^{\parallel\, I} + A_I^\perp P^{\perp\, I}=\tilde A \frac{\Theta_I P^{\parallel \, I}}{\Theta^2} + A_I^\perp P^{\perp\, I}= \tilde A \tilde Q  + A_I^\perp P^{\perp\, I},
\end{aligned}
\end{equation}
where in the last step we defined the gravitino charge 
\begin{equation}
\tilde Q_{ij} = \frac{\Theta^I P_{I\,ij}^\parallel}{\Theta^2}, \qquad P_{I\, ij}^\parallel = \Theta_I \tilde Q_{ij},
\end{equation}
and we implicitly introduced a triplet $\tilde Q^{\rm r}$ such that  $\tilde Q_{ij} = i \tilde Q^{\rm r}\sigma^{r}_{ij}$. 
We can split also the vector fields kinetic term
\begin{equation}
\begin{aligned}
a^{IJ} F_I(A) \wedge * F_J(A) &=a^{IJ} \frac{\Theta_I}{\Theta^2}  \frac{\Theta_J}{\Theta^2} F(\tilde A)\wedge * F(\tilde A) + a^{IJ}  F_I(A^\perp) \wedge * F_J(A^\perp) \\
&=\frac{1}{\Theta^2} F(\tilde A)\wedge * F(\tilde A) +  a^{IJ}  F_I(A^\perp) \wedge * F_J(A^\perp) .
\end{aligned}
\end{equation}
From this expression, we read off the abelian gauge coupling
\begin{equation}
\label{g32N=2}
g^2_{3/2} \equiv \Theta^2 = \Theta_I a^{IJ} \Theta_J
\end{equation}
which, since it enters the gravitino covariant derivative \eqref{gravcovder}, can be understood as the gravitino gauge coupling.
Finally, we look at the scalar potential \eqref{VminimalAdS}, for which we can write 
\begin{equation}
\begin{aligned}
\label{Vads1}
V_{AdS} &= - 2 a^{IJ} P_{I\, ij} P_{J}^{ij} \\
&= - 2a^{IJ} \left(P_{I\, ij}^{\parallel} P_{J}^{\parallel\, ij}+P_{I\, ij}^{\perp} P_{J}^{\perp\, ij}\right)\\
&\leq -2  a^{IJ} P_{I\, ij}^{\parallel} P_{J}^{\parallel\, ij},
\end{aligned}
\end{equation}
where we used the fact that $a_{IJ}$ is positive semi-definite. Inserting the expressions for $g_{3/2}$ and $\tilde Q$, we have
\begin{equation}
V_{AdS}  \leq -2  a^{IJ} P_{I\, ij}^{\parallel} P_{J}^{\parallel\, ij} =  - 2  a^{IJ} \Theta_I \Theta_J \tilde Q_{ij} \tilde Q^{ ij}=- 2 g_{3/2}^2 \tilde Q_{ij} \tilde Q^{ij}
\end{equation}
or in absolute value
\begin{equation}
\label{N=2Vads2}
|V_{AdS}| \geq 2\, g_{3/2}^2 \,\tilde Q_{ij} \tilde Q^{ij}.
\end{equation}

Actually, for maximally supersymmetric anti-de Sitter vacua we can say more. Indeed, one can argue that the prepotentials $P_{I\, ij}^{\perp}$ which have been neglected in the last step of \eqref{Vads1} are in fact vanishing on the vacuum and that formula becomes an equality. By assumption, the $P_{I\, ij}^{\perp}$ should be at most constant, otherwise they would produce non-vanishing Killing vectors and thus break supersymmetry. Being constant, if they do not vanish then they gauge the R-symmetry. More precisely, they gauge the part of $H_R$ which is orthogonal to the one gauged by $P_{I\, ij}^{\parallel}$. However, in our case both the latter and $H_R$ are U(1)$_R$, implying that $P_{I\, ij}^{\perp}$ must vanish on the vacuum, for there is no orthogonal direction to U(1)$_R$ and still inside $H_R= {\rm U(1)}_R$. As a consequences, in the models under investigation the bound \eqref{N=2Vads2} is saturated, namely
\begin{equation}
\label{VadsN=2=}
|V_{AdS}| = 2 g_{3/2}^2 \,\tilde Q_{ij} \tilde Q^{ij} = 4 g_{3/2}^2 \, \tilde Q^{\rm r} \tilde Q^{\rm r} .
\end{equation}
The same argument can be applied to four-dimensional anti-de Sitter vacua with eight preserved supercharges and in particular to the results of \cite{Cribiori:2022trc}.

This shows in a model-independent way that the magnitude of the supersymmetric anti-de Sitter cosmological is coincided with (or at most bounded from below by) the gravitino gauge coupling.
This is valid for any matter content and in particular also for the pure theory with only the gravity multiplet.
Assuming some form of charge quantization, which is anyway expected in quantum gravity, the quantity $\tilde Q^{\rm r} \tilde Q^{\rm r}$ cannot be arbitrarily small.  
Thus, the cosmological constant cannot be decoupled from $g_{3/2}$. From a swampland perspective, this has important consequences on scale separation.

First, it is immediate to see that parametric scale separation is ruled out. Indeed, it would require to take $|V_{AdS}| \to 0$ at fixed Kaluza-Klein scale, but, as a consequence of \eqref{VadsN=2=}, this must be accompanied by $g_{3/2} \to 0$. Assuming it is smooth, such a limit leads to a restoration  of a global (R-)symmetry, something believed not to be consistent in quantum gravity.
This fact has been recently pointed out in \cite{Montero:2022ghl}.
Second, one can argue that the relation \eqref{VadsN=2=} is already problematic as it is, without the need to take any limit. To this purpose, one has to recall that the magnetic weak gravity conjecture postulates that the ultraviolet cutoff $\Lambda_{UV}$ of an effective theory with an abelian gauge symmetry is bounded from above by the gauge coupling $g$ (in Planck units),
\begin{equation}
\Lambda_{UV} \lesssim g .
\end{equation}
Then, we can extend the argument of \cite{Cribiori:2022trc} to five dimensions and apply the weak gravity conjecture to the U(1)$_R$ gauging.\footnote{It is not guaranteed that the weak gravity conjecture, which has been originally formulated in flat space \cite{Arkani-Hamed:2006emk}, extends directly to curved backgrounds. Here, as in \cite{Cribiori:2020use,Cribiori:2022trc}, we assume that the weak gravity conjecture in (anti-)de Sitter receives corrections suppressed with the (anti-)de Sitter radius, for example as argued for in \cite{Huang:2006hc} (see also \cite{Antoniadis:2020xso} for the de Sitter case and \cite{Aharony:2021mpc,Palti:2022unw,Andriolo:2022hax} for another, new, perspective on the weak gravity conjecture in anti-de Sitter). In the limit of small cosmological constant, the radius grows and thus such corrections become negligible.} As a result, we find that the cosmological constant is quantized in terms of the cutoff as
\begin{equation}
\label{VminimalAdSWGC}
|V_{AdS}| = 4\, g_{3/2}^2 \, \tilde Q^{\rm r} \tilde Q^{\rm r} \gtrsim \Lambda_{UV}^2 .
\end{equation}
In supergravity, the ultraviolet cutoff is typically the Kaluza-Klein scale of the compactification. Then, \eqref{VminimalAdSWGC} implies that supersymmetric anti-de Sitter vacua of five-dimensional supergravity cannot be scale separated as a consequence of the weak gravity conjecture, for the cosmological constant and the Kaluza-Klein scale are of the same order. These statements hold regardless of the details of the microscopic description, as long as this admits an anti-de Sitter vacuum with eight supercharges.

\subsubsection{An example: type IIB supergravity on $T^{1,1}$}

Even if the arguments given above are model-independent, it is instructive to see them at work in a specific example. In this respect let us mention that, to the best of our knowledge, no known compactification of critical string theory leads to a five-dimensional supersymmetric anti-de Sitter vacuum with scale separation. The prototype example is in fact type IIB on $AdS_5\times S^5$, where the radii of the two spaces are set to same value by supersymmetry. In the following, we consider another well-known example for illustrative purposes, namely type IIB supergravity compactified on the Sasaki-Einstein manifold $T^{1,1} = {\rm SU}(2)\times {\rm SU}(2)/{\rm U}(1)$ \cite{Cassani:2010na,Bena:2010pr}. 

For simplicity, we consider two related subcases known as Betti-vector and Betti-hyper truncations respectively. 
The original model has $\mathcal{N}=4$ supersymmetry at the lagrangian level, which can be organized into massless and massive $\mathcal{N}=2$ multiplets. 
There are three $\mathcal{N}=4$ vector multiplets, which can be viewed as the union of one $\mathcal{N}=2$ vector and one $\mathcal{N}=2$ hyper multiplet. 
One of these $\mathcal{N}=4$ multiplets is called Betti multiplet, for it arises from dimensional reduction on the 2-form and 3-form which are non-trivial in the cohomology of $T^{1,1}\sim S^2\times S^3$. 
The model can be truncated to $\mathcal{N}=2$ (i.e.~minimal) supersymmetry at the lagrangian level. After truncating the $\mathcal{N}=2$ massive gravitino multiplet, one can achieve a consistent truncation by truncating also the $\mathcal{N}=2$ hyper or the $\mathcal{N}=2$ vector multiplet (or both) constituting the $\mathcal{N}=4$ Betti multiplet. 
In the first case, one ends up with two $\mathcal{N}=2$ vector and two $\mathcal{N}=2$ hyper multiplets. This is called Betti-vector truncation.
In the second case, one ends up with one $\mathcal{N}=2$ vector and three $\mathcal{N}=2$ hyper multiplets. This is called Betti-hyper truncation. 
We analyze both of them below, following mainly \cite{Halmagyi:2011yd,Louis:2016msm} and presenting the relevant quantities in our language.

We start from the Betti-vector truncation. 
This is an $\mathcal{N}=2$ supergravity in five dimensions coupled to $n_V=2$ vector and $n_H=2$ hyper multiplets. 
The whole scalar manifold is given by \eqref{Mscalar} with
\begin{equation}
\mathcal{S} = SO(1,1)\otimes SO(1,1),\qquad  \mathcal{Q}=\frac{SO(4,2)}{SO(4)\times SO(2)}.
\end{equation}
The scalars in the vector multiplets are denoted $u_2, u_3$, while the scalars in the hyper multiplets are $u_1, k$, which are real, and $\tau, b_1, b_2$, which are complex; $\tau$ arises from the type IIB axio-dilaton.  Hence, we have ten scalar fields in total.

As for the scalars in the vector multiplets, we can parametrize them as 
\begin{equation}
h^I=\left(e^{4u_3},e^{2u_2-2u_3},e^{-2u_2-2u_3}\right),\qquad h_I=\frac13\left(e^{-4u_3},e^{-2u_2+2u_3},e^{2u_2+2u_3}\right),
\end{equation}
where $I=0,1,2$ and $C_{012}=\frac16$. 
In this basis, the metric $a_{IJ}$ is diagonal
\begin{equation}
a_{IJ}=\frac13\left(
\begin{array}{ccc}
e^{-8u_3} &0 &0\\
0&e^{-4u_2+4u_3}&0\\
0&0&e^{4u_2+4u_3}
\end{array}\right).
\end{equation}
As for the scalars in the hyper multiplets, we refer to \cite{Louis:2016qca,Halmagyi:2011yd} with the only difference that, in order to simplify the analysis, here we set $b_1=b_2=0$ from the beginning, since in any case it is going to hold in the vacuum. Indeed, we do not want to re-derive a known result, rather we want to interpret it from our perspective.

In this model, the gauging is performed solely on the quaternionic manifold. The non-vanishing Killing vectors are along the direction $u=k$ and read
\begin{align}
K_0^k=-\xi\partial_k, \qquad K_1^k=2 \partial_k, \qquad K_2^k=2 \partial_k,
\end{align}
 where $\xi>0$ is a parameter. The associated quaternionic prepotentials are
 \begin{align}
 P_0^{\rm r} = \left(\frac{\xi}{2} e^{-4 u_1}-3\right)\delta^{{\rm r}3},\qquad P_1^{\rm r} = - e^{-4 u_1}\delta^{{\rm r}3},\qquad P_2^{\rm r} = - e^{-4 u_1}\delta^{{\rm r}3}.
 \end{align}
 The conditions \eqref{susyAdSN=2} for a supersymmetric anti-de Sitter vacuum fix the scalars
 \begin{equation}
 u_1 = -\frac32 u_3,\qquad u_2=0,\qquad e^{-6u_3}=\frac {\xi}{4},
 \end{equation}
 and the cosmological constant is
 \begin{equation}
 \label{ccBettivec}
 |V_{AdS}| = 36 \left(\frac{4}{\xi}\right)^{\frac 43}.
 \end{equation}
 Given that we set already $b_1=b_2=0$, the fields $k$ and $\tau$ remain unfixed. The former is a goldstone boson, while the latter is a complex modulus of the model. The moduli space has thus real dimension two. 

From the general analysis of \cite{Lust:2017aqj}, we expect an unbroken $H_R = {\rm U(1)}_R$ group in the vacuum. The associated prepotential can be identified by constructing a linear combination of $P_I^{\rm r}$ such that any field dependence drops out. Indeed, this identifies a unique combination up to an overall factor, which can then be fixed by matching with the vacuum energy. We thus find that the unbroken $U(1)_R$ gauge group is associated to a constant prepotential $\Theta^I P_I^{\rm r}$, where
\begin{equation}
\Theta^I =3\left(\frac{4}{\xi}\right)^\frac43 \left(\tilde Q^{\rm r}\tilde Q^{\rm r}\right)^{-\frac12}\left(1,\frac{\xi}{4},\frac{\xi}{4}\right).
\end{equation}
The vector gauging this group is then $\tilde A = \Theta_I A^I$, where $\Theta_I = a_{IJ}\Theta^J$. As a consistency check, one can notice that $\Theta^I K_I^{k}=0$, confirming that this is indeed an R-symmetry. Having identified the coefficients $\Theta^I$, we can now calculate the gravitino gauge coupling \eqref{g32N=2} and we find that
\begin{equation}
\label{g32Bettivec}
g_{3/2}^2 \tilde Q^{\rm r}\tilde Q^{\rm r} = 9 \left(\frac{4}{\xi}\right)^\frac{4}{3}.
\end{equation}
By combing this with the vacuum energy \eqref{ccBettivec}, we recover the general expression \eqref{VadsN=2=}. For the arguments given in section \ref{sec_noscalesepN=2}, we conclude that these vacua are not scale separated.

Next, we analyze the Betti-hyper truncation. This is an $\mathcal{N}=2$ supergravity in five dimensions coupled to $n_V=1$ vector and $n_H=3$ hyper multiplets. 
The whole scalar manifold is given by \eqref{Mscalar} with
\begin{equation}
\mathcal{S} = SO(1,1),\qquad  \mathcal{Q}=\frac{SO(4,3)}{SO(4)\times SO(3)}.
\end{equation}
The scalar in the vector multiplet is denoted with $u_3$, while the scalars in the hyper multiplets are $u_1, k, e_1, e_2$, which are real, and $\tau,v, b_1, b_2$, which are complex; $\tau$ arises from the type IIB axio-dilaton.  Hence, we have thirteen scalar fields in total.

As for the scalar in the vector multiplet, we can parametrize it as 
\begin{equation}
h^I=\left(e^{4u_3},e^{-2u_3}\right),\qquad h_I=\frac13\left(e^{-4u_3},2e^{2u_3}\right),
\end{equation}
where $I=0,1,$ and $C_{011}=\frac13$. 
In this basis, the metric $a_{IJ}$ is diagonal
\begin{equation}
a_{IJ}=\frac13\left(
\begin{array}{cc}
e^{-8u_3} &0 \\
0&2e^{4u_3}\\
\end{array}\right).
\end{equation}
As for the scalars in the hyper multiplets, we refer to \cite{Louis:2016qca,Halmagyi:2011yd} with the only difference that, in order to simplify the analysis, here we set $b_1=b_2=0$ from the beginning, since in any case it is going to hold in the vacuum. For the same reason, we also set $j^1=j^2=0$, $j^1,j^2$ being the charges associated to $e^1, e^2$.

In this model, the gauging is performed solely on the quaternionic manifold. The non-vanishing Killing vectors are along $u=k,\rho$, where $\rho$ is a complex scalar depending on $v,\bar v$ \cite{Halmagyi:2011yd,Louis:2016qca}; they read
\begin{align}
\Vec{K}_0 = - \xi\partial_k + \frac32 (1+ \rho^2) \partial_\rho + \frac32 (1+ \bar \rho^2)\partial_{\bar \rho},\qquad \Vec{K}_1 =4\partial_k,
\end{align}
 where $\xi>0$ is a parameter. The associated quaternionic prepotentials are
\begin{align}
P_0^{\rm r} = \left(\frac \xi2 e^{-4u_1} - \frac{3}{2{\rm Im}\rho}(1+|\rho|^2)\right)\delta^{{\rm r}3},\qquad P_1^{\rm r} = -2 e^{-4u_1}\delta^{{\rm r}3}.
\end{align}
The conditions \eqref{susyAdSN=2} for a supersymmetric anti-de Sitter vacuum fix the scalars
\begin{equation}
u_1 = -\frac32 u_3,\qquad e^{-6u_3}=\frac {\xi}{4}, \qquad \rho=i,
\end{equation}
and the cosmological constant is given again by \eqref{ccBettivec}. 
Since we have already set $b_1=b_2=0$, the fields $k,\tau,e_1,e_2$ remain unfixed. The former is a goldstone boson, while the latter are moduli of the model. The moduli space has thus real dimension four. 

Proceeding as for the previous example, we identify the linear combination of prepotentials $\Theta^I P_I^{\rm r}$ associated the gauging of $H_R={\rm U(1)}_R$. This is given by the coefficients
\begin{equation}
\Theta^I =3\left(\frac{4}{\xi}\right)^\frac43 \left(\tilde Q^{\rm r}\tilde Q^{\rm r}\right)^{-\frac12}\left(1,\frac{\xi}{4}\right)
\end{equation}
and thus the corresponding vector field is $\tilde A = \Theta_I A^I$, with $\Theta_I = a_{IJ}\Theta^J$. As a consistency check, one can see once more that $\Theta^I K_I^{u}=0$ on the vacuum. The gravitino gauge coupling \eqref{g32N=2} is again \eqref{g32Bettivec} and, by combining it with the vacuum energy, we recover the general expression \eqref{VadsN=2=}. For the arguments given in section \ref{sec_noscalesepN=2}, we conclude that these vacua are not scale separated.

\subsection{Scale separation, holography and the species scale}
\label{sec:holspecies}

Since we are considering supersymmetric anti-de Sitter vacua, we can also provide a complementary argument against scale separation based on holography and on the concept of species scale.
This is independent from any of the swampland conjectures mentioned so far. A more concrete version of such an argument has been used in \cite{Lust:2022lfc} to show that KKLT-like anti-de Sitter vacua in three and four dimensions cannot feature an arbitrary small (in magnitude) cosmological constant. Here, we are mainly interested in the five-dimensional case.

As originally proposed in \cite{Dvali:2007wp,Dvali:2007hz,Dvali:2010vm}, the ultraviolet cutoff of a $d$-dimensional effective theory of gravity coupled to a number $N_{sp}$ of light species is not the Planck mass $M_P$, but rather the so-called species scale
\begin{equation}
\Lambda_{sp} \simeq \frac{M_P}{N_{sp}^{\frac{1}{d-2}}}.
\end{equation}
In general, it can be difficult to give an estimate of $N_{sp}$.\footnote{A recent proposal for a moduli-dependent definition of $N_{sp}$ on the vector multiplet moduli space of $\mathcal{N}=2$ supergravity in four dimensions has been given in \cite{vandeHeisteeg:2022btw} and motivated from black hole physics in \cite{Cribiori:2022nke}.} 
However, it is reasonable to assume that the number of light species in the five-dimensional anti-de Sitter theory and in the dual four-dimensional conformal field theory be equal. An estimate for the latter is given by the central charge $a$, namely the coefficient of the Euler density in the Weyl anomaly \cite{vandeHeisteeg:2022btw}
\begin{equation}
a \simeq N_{sp}.
\end{equation}
For five-dimensional anti-de Sitter vacua with holographic dual, the cosmological constant is also given in terms of $a$ as $M_P^{-2}\,|V_{ AdS}| \simeq a^{-\frac23}$ and thus
\begin{equation}
M_P^{-2}\,|V_{ AdS}| \simeq a^{-\frac23} \simeq (N_{sp})^{-\frac23},
\end{equation}
leading in turn to
\begin{equation}
N_{sp} \simeq \left(\frac{M_P}{\mathbb{H}}\right)^3,
\end{equation}
where we introduced the Hubble parameter $\mathbb{H}$, such that $|V_{AdS}|=6\mathbb{H}^2$.
Hence, in such five-dimensional gravitational theories, the species scale is in fact
\begin{equation}
\label{Lambdasp}
\Lambda_{sp} \simeq \frac{M_P}{N_{sp}^{\frac13}} \simeq \sqrt{|V_{AdS}|} \simeq \mathbb{H}.
\end{equation}
Notice that the explicit dependence on the Planck mass drops out of this relation. 

At this point, in order to say something about scale separation, we need to specify how the Kaluza-Klein scale is related to $\Lambda_{sp}$. If these two scales are one and the same, then \eqref{Lambdasp} forbids scale separation in these vacua. If instead the Kaluza-Klein scale is bigger than the species scale, we cannot draw the same conclusion. In any case, the relation \eqref{Lambdasp} tells us that the limit of parametrically small (in magnitude) cosmological constant leads to a large number of light species in the effective theory, in accordance with \cite{Lust:2019zwm}.

The above argument can be in fact formulated in arbitrary dimensions. The central charge of a $(d-1)$-dimensional conformal field theory is given in terms of the Hubble scale of the dual $d$-dimensional theory of gravity as $c \simeq \left(M_P/\mathbb{H}\right)^{d-2}$. Since for conformal field theories with an holographic dual one must have $c=a$, one can speculate that $c \simeq N_{sp}$ in generic dimensions. Assuming this speculation to be legitimate and repeating the same steps as above, one arrives again at the relation $\Lambda_{sp} \simeq \mathbb{H}$. 
Anti-de Sitter vacua in which the Kaluza-Klein scale and the species scale coincide cannot therefore be scale separated. 
In case this turns out not to be the case for the DGKT class of vacua \cite{DeWolfe:2005uu} (assuming their holographic dual exists), we can still notice that their regime of parametric control leads to a parametrically small species scale which usually signals a breakdown of the effective description.
Given the relevance of this matter for the problem of scale separation, it would be important to make these statement more concrete; we hope to come back to them in the future.

\subsection{No effective description in de Sitter}
\label{sec:nodSN=2}

We consider now de Sitter vacua and show how swampland conjectures can be used to constraint them. 
As it will be clear, to give constraints on de Sitter vacua is more challenging than to restrict scale separation on supersymmetric anti-de Sitter vacua. 
Indeed, while in anti-de Sitter we have been able to give a general model-independent argument, in de Sitter we have either to introduce more assumptions or to perform a case by case analysis. 
In this section, we will first generalize to five dimensions the argument of \cite{Cribiori:2020use,DallAgata:2021nnr,Cribiori:2022sxf} ruling out the existence of de Sitter vacua with parametrically small gravitino mass, and then we will comment on the models which evade such an assumption. Unfortunately, the few stable de Sitter vacua known in the literature \cite{Cosemans:2005sj, Ogetbil:2008tk,Ogetbil:2008ojt} are of this kind and thus, contrary to the four-dimensional case, we will not draw any conclusion on them.

The starting point is the scalar potential \eqref{VN=2gen}, where we set $g_R =g=1$ as usual. Assuming that the gravitino mass is vanishing on the vacuum, $P_{ij}P^{ij}= (h^I P_{I\,ij})( h^J P_J^{ij})=0$, the potential reduces to
\begin{equation}
\begin{aligned}
V_{dS}&= \left(P^{\tilde x}_{ij}P^{ij}_{\tilde x} - 2P_{ij}P^{ij}\right) + 2W^{\tilde x} W_{\tilde x}+2\mathcal{N}_{iA}\mathcal{N}^{iA},
\end{aligned}
\end{equation}
where we used that $P^{\tilde x}_{ij} P^{ij}_{\tilde x} = a^{IJ} P_{I\,ij} P_{J}^{ij}$ if $h^I P_{I\,ij}=0$. This scalar potential is manifestly positive semi-definite
\begin{equation}
V_{dS} =a^{IJ} P_{I\,ij} P_{J}^{ij} + 2W^{\tilde x} W_{\tilde x}+2\mathcal{N}_{iA}\mathcal{N}^{iA}
\end{equation}
and since every term in the sum is positive, we can write
\begin{equation}
\label{VminimaldS}
V_{dS} \geq  a^{IJ} P_{I\,ij} P_{J}^{ij}.
\end{equation}
The same conclusion can be reached even if the gravitino mass is not exactly vanishing but still parametrically small compared to the vacuum energy, namely $P_{ij}P^{ij} / V_{dS}\ll 1$. Thus, we directly extend our discussion to this case.

The result \eqref{VminimaldS} is analogous to \eqref{VminimalAdS} for the anti-de Sitter case: the quantity on the right hand side is the product of the gauge coupling and the charge without any additional arbitrary parameter entering the relation. 
To see this more explicitly, we can employ the same field redefinitions of the previous section to show that
\begin{equation}
\begin{aligned}
a^{IJ} P_{Iij}P^{ij}_J \geq a^{IJ} P^\parallel_{Iij}P^{\parallel\, ij}_J = g_{3/2}^2 \tilde Q_{ij} \tilde Q^{ij} 
\end{aligned}
\end{equation}
and thus 
\begin{equation}
\label{VminimaldS2}
V_{dS} \geq  g_{3/2}^2 \tilde Q_{ij} \tilde Q^{ij} .
\end{equation}

This formula is the starting point to argue that such vacua are not trustable effective field theory. First, we can see that, similarly to the anti-de Sitter case, we cannot have a parametrically small cosmological constant, unless $g_{3/2}\to 0$. However, assuming it is smooth, this limit leads to the restoration of a global (R-)symmetry, which is not compatible with quantum gravity. Notice that the same conclusion can be reached in four dimensions by adapting this reasoning to the results of \cite{Cribiori:2020use,DallAgata:2021nnr}.

Nevertheless, for finite $g_{3/2}$ one could hope that these vacua are trustworthy. 
That this is not the case can be seen following \cite{Cribiori:2020use,DallAgata:2021nnr} and using the weak gravity conjecture. 
At this point, it is important to recall that the Hubble scale on a de Sitter background gives a natural proxy for the infrared cutoff of the theory, $\mathbb{H}\sim \Lambda_{IR}$, for the only meaningful distances that one can probe are bounded by the de Sitter horizon radius. On the other hand, the magnetic weak gravity conjecture postulates that the ultraviolet cutoff is given by the (abelian) gauge coupling. Thus, the relation \eqref{VminimaldS2} is telling us that
\begin{equation}
\Lambda_{IR}^2\sim \mathbb{H}^2 \simeq V_{dS} \geq g_{3/2}^2 \tilde Q_{ij} \tilde Q^{ij}  \gtrsim \Lambda_{UV}^2,
\end{equation}
namely the infrared and the ultraviolet cutoff are roughly of the same order. As a consequence, these models cannot be trustworthy effective theories since they are receiving any sort of correction.

Notice that we are using the weak gravity conjecture in the same spirit as in the anti-de Sitter case: we assume that corrections to the flat space formulation are negligible at large Hubble radius. However, an important difference with respect to the anti-de Sitter case is that we are now not guaranteed that an abelian factor of the gauge group is unbroken in the vacuum. In case the vacuum preserves a non-abelain gauge symmetry, we can apply the weak gravity conjecture with respect to any of the vector fields gauging the Cartan subalgebra. We will come back to this when discussing scale separation in maximal supergravity.

Interestingly, the fact that parametrically light gravitini lead to a breakdown of the effective field theory in de Sitter is in accordance with the bound of \cite{Montero:2019ekk,Montero:2021otb}, which forbids massless charged particles in a consistent theory of quantum gravity on a positive background. Indeed, here the gravitini are charged, for their gauge coupling $g_{3/2}$ is non-vanishing on the vacuum.

\subsubsection{A class of unstable vacua}

We present a class of models for which our argument can be run and thus the associated de Sitter critical points are in tension with the weak gravity conjecture. This class is taken from \cite{Gunaydin:2000xk,Cosemans:2005sj} and its scalar potential is generated by a U(1)$_R$ gauging with no hyper multiplets but with non-vanishing $P^{\tilde x\,\rm r}$, namely
\begin{equation}
V = 2 P^{\tilde x\, {\rm r}} P_{\tilde x}^{\rm r} - 4 P^{\rm r} P^{\rm r}=-4 C^{IJ\tilde K}h_{\tilde K}P_I^{\rm r}P_J^{\rm r},
\end{equation}
where we used \eqref{aij=hh} and \eqref{aijdef}. Assuming also that no tensor multiplets are contributing, the potential simplifies to
\begin{equation}
V=-4 C^{IJ K}h_{K}P_I^{\rm r}P_J^{\rm r}
\end{equation}
and the scalar dependence is only in $C^{IJK}$ and in $h_K$, the  $P_I^{\rm r}$ being constant. Critical points are solutions of 
\begin{equation}
\label{dxVexampledS}
\partial_x V = -4 \left(\partial_x C^{IJK} h_KP_I^{\rm r}P_J^{\rm r}+C^{IJK}\partial_x h_KP_I^{\rm r}P_J^{\rm r}\right)=0.
\end{equation}
However, one can show that \cite{Cosemans:2005sj}
\begin{equation}
\partial_u C^{IJK} = 2 h^{Ix} h^{Jy} h^{Kz}D_uT_{xyz} ,\qquad D_uT_{xyz} = \sqrt{\frac32} \left(g_{(xy}g_{z)u} - 2 {T_{(xy}}^w T_{z)uw}\right),
\end{equation}
where $T_{xyz}=h^{Ix} h^{Jy} h^{Kz} C_{IJK}$ and therefore the first term on the right hand side of \eqref{dxVexampledS} is actually vanishing as a consequence of \eqref{hix}. Then, by contracting \eqref{dxVexampledS} with $h^I_x$ and using  \eqref{aij=hh}, \eqref{aijdef} we find after few steps
\begin{equation}
\label{CPP=Vh}
C^{IJK}P^{\rm r}_I P^{\rm r}_J=-\frac14 V h^K.
\end{equation}
Thus the problem of finding critical points has become algebraic.

From this relation, one can prove that very special manifolds associated to Jordan algebras cannot give de Sitter critical points when only a U(1)$_R$ gauging is turned on \cite{Gunaydin:1984ak}. Let us denote $\hat P^I = C^{IJK}P^{\rm r}_I P^{\rm r}_J$. Then, by exploiting properties of Jordan algebras, one can show that \cite{Gunaydin:1983bi}
\begin{equation}
C_{IJK} \hat P^I \hat P^J \hat P^K = \left(C_{IJK}P^{I \rm r}P^{J \rm r} P^{K \rm s}\right)^2>0.
\end{equation}
On the other hand, from \eqref{CPP=Vh} we have
\begin{equation}
C_{IJK} \hat P^I \hat P^J \hat P^K = \left(-\frac14 V\right)^3.
\end{equation}
By comparing these last two relations we see that if $V >0$ at the critical point we are led to a contradiction. 
Thus, to find de Sitter critical points one has to deviate from Jordan algebras.

In \cite{Cosemans:2005sj}, de Sitter critical points with only U(1)$_R$ gauging are found starting from
\begin{equation}
\mathcal{F}(h) = (h^0)^3 - \frac32 h^0 h^a h^b \delta_{ab}+C_{abc}h^ah^bh^c,
\end{equation}
where we split $I=(0,a)$.
For this model, the equation \eqref{CPP=Vh} gives two conditions
\begin{align}
&C^{0JK} P_J^{\rm r} P_K^{\rm r} = -\frac14 V h^0,\\
&C^{abc} P_b^{\rm r} P_c^{\rm r} +2C^{a0b}P_0^{\rm r} P_b^{\rm r} =0.
\end{align}
One can always perform a linear transformations of the $h^I$ such that at the critical point $h^I=(h^0,h^a)=(1,0)$ and $a_{IJ}=\delta_{IJ}$, implying in turn $C_{IJK}=C^{IJK}$ at the same point. Performing this step, the equations above become
\begin{align}
\label{dSU1eq1}
&C^{0JK} P_J^{\rm r} P_K^{\rm r} = -\frac14 V ,\\
\label{dSU1eq2}
&C^{abc} P_b^{\rm r} P_c^{\rm r}  +2C^{a0b}P_0^{\rm r} P_b^{\rm r}=  0,
\end{align}
and the vacuum energy is then 
\begin{equation}
V   = -4 P_0^{\rm r} P_0^{\rm r} + 2P_a^{\rm r} P_b^{\rm r}\delta^{ab} . 
\end{equation}
Thus, to have de Sitter we have to find solutions of these equations such that
\begin{equation}
-4 P_0^{\rm r} P_0^{\rm r} + 2P_a^{\rm r} P_b^{\rm r}\delta^{ab} >0
\end{equation} 
To solve \eqref{dSU1eq2} for generic $C_{abc}$ is not straightforward. Indeed, \cite{Cosemans:2005sj} propose to set $C_{abc}=0$ and then there are two solutions: either $P_a^{\rm r}=0$ or $P_0^{\rm r}=0$. The first solution corresponds to anti-de Sitter critical points, while the second to de Sitter. Crucially, for the latter one has $h^I P_I^{\rm r} =0$ and thus the gravitino mass is necessarily vanishing. Therefore, these de Sitter critical points are not trustworthy effective field theories according to the weak gravity conjecture.

Notice that we do not have to analyze stability when applying our argument: we can draw conclusions just by looking at the contributions to the vacuum energy. In any case in \cite{Cosemans:2005sj} it is proved that, whenever they exists, de Sitter critical points like these which are stemming solely from a U(1)$_R$ gauging are at best unstable.

\subsubsection{On the fate of stable vacua}

In the work \cite{Cosemans:2005sj}, the first examples of stable de Sitter vacua of minimal supergravity in five dimensions are provided. We would like to check the gravitino mass on them, to see whether or not our argument can apply.

The class of models contains both vectors and tensor multiplets, but no hyper multiplets, and the potential stems from an SO$(1,1)\times$U$(1)_R$ gauging. The very special geometry is of the generic Jordan type and fixed by
\begin{equation}
\mathcal{F}(h) = 3 \frac{\sqrt 3}{2} h^0\left(h^a \eta_{ab}h^b\right), \qquad \eta_{ab} = \text{diag}(1,-1,-1,\dots).
\end{equation}
The constraint $\mathcal{F}(h)\equiv 1$ is solved by
\begin{equation}
\label{hphiJord}
h^0=\frac{1}{\sqrt 3}\frac{1}{||\phi||^2}, \qquad h^a= \sqrt{\frac23}\phi^a, \qquad ||\phi||^2=\phi^a\eta_{ab}\phi^b
\end{equation}
and we restrict to the region $||\phi||^2>0$, $\phi^1>0$.
The presence of tensor fields gives an additional contribution to the scalar potential which we did not mention in section \ref{sec:revminsugra}. Below, we just report the main results from \cite{Cosemans:2005sj} and we refer to it for more details. 

Without loss of generality, we can rotate the SU(2)$_R$ frame in such a way that only one of the $P^{\rm r}_I$ is non-vanishing. Thus, we drop the index $\rm r=1,2,3$ in what follows. A direct calculation shows that at the critical point the physical scalar fields are fixed by
\begin{align}
\label{stabledSCosSmet}
&\frac{\phi^a}{||\phi||^4} = 16 \sqrt{2} P_0 P_a,\\
&||\phi||^{-6} = -\frac12 \left(16 \sqrt 2 P_0 P_a P_b \delta^{ab}\right)^2 +8 P_a P_b \delta^{ab}.
\end{align}
The vacuum energy is
\begin{equation}
V= 3||\phi||^2 P_a P_b\delta^{ab}\left(1-32 P_0P_0\right)
\end{equation}
and it is positive for $32 P_0 P_0<1$. One can then check that these de Sitter critical points are indeed stable \cite{Cosemans:2005sj}.

We can see that it is not possible to set the gravitino mass to zero on these vacua. Using the parametrization \eqref{hphiJord} for the generic Jordan family and imposing
\begin{equation}
h^I P_I^r = \frac{1}{\sqrt 3} \frac{1}{||\phi||^2} P_0^{\rm r} + \sqrt{\frac23} \phi^a P_a^{\rm r} =0,
\end{equation}
we have
\begin{equation}
P_0^{\rm r} = -\sqrt2 ||\phi||^2 \phi^a P_a^{\rm r}.
\end{equation}
Substituting now the values of the scalars at the critical point \eqref{stabledSCosSmet}, we get $P_0 = -32 P_0 P_a P_b\delta^{ab}/||\phi||^2$, which has unique solution $P_0=0$. However, this would imply $\phi^a=0=||\phi||^2$ and thus the vacuum energy would vanish. Hence, these de Sitter vacua necessarily require a non-vanishing gravitino mass. An analogous check on (some of) the stable models in \cite{Ogetbil:2008ojt,Ogetbil:2008tk} led us to a similar outcome. Given that the gravitino mass is non-vanishing, we cannot apply our general argument. Other tools are needed and we leave this investigation for future work. The most we can say at present is that a parametrically small gravitino mass obtained with $P_0^{\rm r}\to 0$ is problematic, for it would lead to $V\sim g_{3/2}^2\to 0$ and thus to the restoration of a global (R-)symmetry.\footnote{That the limit of parametrically small gravitino mass leads to the breakdown of the effective description has been formulated as a swampland conjecture in \cite{Cribiori:2021gbf,Castellano:2021yye}; see also \cite{Antoniadis:1988jn}.}
Indeed, in the present class of models the parameter  $P_0^{\rm r}$ governs both the cosmological constant and the gravitino mass.

We conclude by noticing that this situation is somehow different from its lower dimensional counterpart. Indeed, all known stable de Sitter vacua of $\mathcal{N}=2$ supergravity in four dimensions feature a vanishing gravitino mass and are thus in tension with the weak gravity conjecture \cite{Cribiori:2020use,DallAgata:2021nnr}.
Given that swampland conjectures are typically well-behaved under dimensional reduction\footnote{
To the best of our knowledge, there is no known example in which a dimensionally reduced theory does not satisfies the sub-lattice weak gravity conjecture whereas the parent one does \cite{Heidenreich:2015nta}.}, this suggest that the five-dimensional stable de Sitter vacua are somehow different from the known four-dimensional ones. As such, they deserve further investigation which we leave for the future.

\section{Constraints on maximal supergravity}

We want now to extend the analysis to maximal supergravity in five dimensions. To this purpose, we first review some elements of the gauged theory, following notation and conventions of \cite{deWit:2004nw}, to which we refer for more details. Then, in analogy to what done for the minimal theory, we study constraints on scale separation and on de Sitter vacua.

\subsection{Maximal supergravity in five dimensions}

Due to maximal supersymmetry, all fields are part of the gravity multiplet, which contains: the graviton, 8 gravitini $\psi^i_\mu$, with $i=1,\dots,8$ fundamental index of the USp$(8)$ R-symmetry, 27 vector fields $A_\mu^{M}$, with $M=1,\dots,27$ anti-fundamental (\bm{$\overline{27}$}) index of $E_{6(6)}$, 48 spin-1/2 fields $\chi^{ijk}$ and 48 scalars ${\mathcal{V}_M}^{ij}$, with $M$ now fundamental ({\bf 27}) index of $E_{6(6)}$. The scalar fields parametrize the coset
\begin{equation}
\frac{G}{H} = \frac{E_{6(6)}}{{\rm USp}(8)},
\end{equation}
where the R-symmetry group $H={\rm USp}(8)$ is the maximal compact subgroup of $G=E_{6(6)}$. The matrix $\mathcal{V}_M^{ij}=\mathcal{V}_M^{[ij]}$ encoding the scalar fields is pseudoreal and symplectic-traceless,
\begin{equation}
\mathcal{V}_{M\,ij}\equiv({\mathcal{V}_M}^{ij})^* = {\mathcal{V_M}}^{kl}\Omega_{ki}\Omega_{lj},\qquad {\mathcal{V}_M}^{ij} \Omega_{ij}=0,
\end{equation}
where we denoted with $\Omega_{ij}$ the symplectic metric, such that $\Omega_{ij} = -\Omega_{ji}$ and $\Omega_{ij}\Omega^{kj}=\delta_k^j$. The inverse matrix ${\mathcal{V}_{ij}}^M$ is such that
\begin{align}
{\mathcal{V}_M}^{ij}{\mathcal{V}_{ij}}^N = {\delta_M}^N,\qquad {\mathcal{V}_M}^{kl}{\mathcal{V}_{ij}}^M = {\delta_{ij}}^{kl} - \frac18 \Omega_{ij}\Omega^{kl}.
\end{align}

The bosonic Lagrangian is given by
\begin{equation}
\begin{aligned}
e^{-1}\mathcal{L}= -\frac12 R - \frac{1}{16}\mathcal{M}_{MN}{\mathcal{H}_{\mu\nu}}^{M} {\mathcal{H}^{\mu\nu\,N}} -\frac{1}{12}|{\mathcal{P}_{\mu}}^{ijkl}|^2+ \frac{\sqrt 5}{32 e} \mathcal{L}_{VT} -  V.
\end{aligned}
\end{equation}
 The objects
\begin{equation}
{\mathcal{H}_{\mu\nu}}^M ={{F}_{\mu\nu}}^M + g Z^{MN} B_{\mu \nu \, N},
\end{equation}
are a combination of the vector field strengths, ${{F}_{\mu\nu}}^M =\partial_\mu {A_\nu}^M-\partial_\nu {A_\mu}^M-g {f_{NP}}^M {A_\mu}^N{A_\nu}^P$, and tensors intertwined by $Z^{MN}$, which is anti-symmetric tensor transforming in the \bm{$351$} of $E_{6(6)}$.\footnote{As mentioned when discussing the minimal theory, in five dimensions the duality between vectors and tensors is broken by the gauging. To maintain manifest $E_{6(6)}$ covariance, the approach of \cite{deWit:2004nw} is to introduce from the beginning \bm{$\overline{27}$} vectors together with  \bm{$27$} tensors. Then, in order not to increase the number of bosonic degrees of freedom, these fields have to be related by an additional gauge invariance. The gauge invariant object $Z^{MN}$, whose existence follows from the constraints on the embedding tensor (indeed such constraints force the embedding tensor to transform in the \bm{$351$}), allows one to remove vectors in terms of tensors and vice versa. In this manner, vectors and tensor are treated on equal footing and all details of the gauging are delegated to the embedding tensor, or equivalently to $Z^{MN}$.}
Here, ${f_{NP}}^M$ are the structure constants of the gauge algebra and, as for the minimal theory, we introduced an non-physical parameter $g$ to keep track of the gauging deformation.
The gauge kinetic matrix is
\begin{equation}
\label{MMNkin}
\mathcal{M}_{MN} = {\mathcal{V}_M}^{ij}{\mathcal{V}_N}^{kl}\Omega_{ik}\Omega_{jl}=  {\mathcal{V}_M}^{ij}{\mathcal{V}_N}_{ij}
\end{equation}
and is positive definite.  
The object
\begin{equation}
{\mathcal{P}_\mu}^{ijkl} = {\mathcal{V}_{mn}}^M \partial_\mu {\mathcal{V}_M}^{[ij} \Omega^{k|m|}\Omega^{l]n} - g {A_\mu}^M {\mathcal{P}_M}^{ijkl}
\end{equation}
is the USp$(8)$-covariant tensor building the kinetic terms for the scalars and ${\mathcal{P}_M}^{ijkl}$ are the Killing vectors of $E_{6(6)}/{\rm USp}(8)$ coupling to the vector fields ${A_\mu}^M$. 
The explicit expression of $\mathcal{L}_{VT}$ can be found \cite{deWit:2004nw}, but it is not going to be needed in what follows.
The scalar potential is given by
\begin{equation}
V =g^2\left(\frac13 |{A_2}^{i,jkl}|^2-3|{A_1}^{ij}|^2\right),
\end{equation}
where ${A_2}^{i,jkl}$ and ${A_1}^{ij}$ are pseudoreal, symplectic traceless tensors and satisfy $A_1^{[ij]}=0={A_2}^{[i,jkl]}$, ${A_2}^{i,jkl} = {A_2}^{i,[jkl]}$. They are associated to the branching of the embedding tensor under USp$(8)$, $\bm{351}\to\bm{36}+\bm{315}$, and enter the supersymmetry transformations of the fermionic fields
\begin{align}
\delta \psi_\mu^i &= D_\mu \epsilon^i + i \left[\frac{1}{12}\left(\gamma_{\mu\nu\rho} \mathcal{H}^{\nu\rho\,ij}-4 \gamma^\nu {\mathcal{H}_{\mu\nu}}^{ij}\right)-g \gamma_\mu {A_1}^{ij}\right]\Omega_{jk}\epsilon^k,\\
\delta \chi^{ijk} &=\frac i2 \gamma^\mu {\mathcal{P}_\mu}^{ijkl} \Omega_{lm}\epsilon^m - \frac{3}{16} \gamma^{\mu\nu} \left[{\mathcal{H}_{\mu\nu}}^{[ij} \epsilon^{k]}-\frac13 \Omega^{[ij}{\mathcal{H}_{\mu\nu}}^{k]m}\Omega_{mn}\epsilon^n\right] \\
\nonumber
&+ g {A_2}^{l,ijk} \Omega_{lm}\epsilon^m.
\end{align}
Notice that we introduced the notation $|{A_1}^{ij}|^2 \equiv A_{1\,ij} ({A_{1\,ij}})^* ={A_{1\,ij}} {A_1}^{ij}$ and similarly $|A_{2\,i,jkl}|^2 \equiv A_{2\,i,jkl} (A_{2\,i,jkl})^* = A_{2\,i,jkl} {{A_2}^{i,jkl}}$. 
As one can see, ${A_1}^{ij}$ is the gravitino mass matrix, while ${A_2}^{i,jkl}$ encodes the shifts in the supersymmetry transformation of the spin-1/2 fermions. 
The only part of the fermionic couplings that we need to recall is the gravitino covariant derivative, which reads
\begin{equation}
\label{gravcovderN=8}
D_\mu{\psi_\nu}^i = \partial_\mu \psi_\nu^i - \frac14 {\omega_\mu}^{ab}\gamma_{ab}\psi_\nu^i - {\mathcal{Q}_{\mu\,j}}^i \psi_\nu^j,
\end{equation}
where
\begin{align}
{\mathcal{Q}_{\mu\, i}}^j &= \frac13 {\mathcal{V}_{ik}}^N \, \partial_
\mu{\mathcal{V}_{M}}^{jk} - g {A_\mu}^M{\mathcal{Q}_{M\, i}}^j
\end{align}
is the composite USp$(8)$ connection with gauging contribution $ {\mathcal{Q}_{M\, i}}^j$.

For the purposes of our analysis, it is convenient to rewrite the scalar potential in terms of the Killing vectors ${\mathcal{P}_M}^{ijkl}$ and of the gauging couplings $ {\mathcal{Q}_{M\, i}}^j$.  As explained in the appendix \ref{app_N=8}, one can related these quantities to the tensors  ${A_1}^{ij}$ and  ${A_2}^{i,jkl}$ as 
\begin{align}
\label{Q2fin}
\mathcal{M}^{MN} {\mathcal{Q}_{M\, i}}^j {\mathcal{Q}_{N\, j}}^i 
&=-\frac23|{A_{2\,i,jkl}}|^2-\frac{15}{2} \,|{A_{1\,ij}}|^2,\\
\label{P2fin}
\mathcal{M}^{MN}{\mathcal{P}_{M\,ijkl}}{\mathcal{P}_{M}}^{ijkl} &= 6 |A_{2\,i,jkl}|^2.
\end{align}
Thus, the scalar potential becomes
\begin{equation}
\begin{aligned}
\label{VP2Q2}
V &= \frac13 |{A_{2\,i,jkl}}|^2 - 3|{A_{1\,ij}}|^2\\
&=\frac{1}{10} \mathcal{M}^{MN}{\mathcal{P}_{M\,ijkl}}{\mathcal{P}_{M}}^{ijkl} + \frac25 \mathcal{M}^{MN} {\mathcal{Q}_{M\, i}}^j {\mathcal{Q}_{N\, j}}^i.
\end{aligned}
\end{equation}

\subsection{No scale separation in anti-de Sitter}

On a maximally symmetric vacuum, the conditions for unbroken supersymmetry reduce to 
\begin{align}
0&=\delta \psi_\mu^i = D_\mu \epsilon^i - i g \gamma_\mu {A_1}^{ij}\Omega_{jk}\epsilon^k,\\
0&=\delta \chi^{ijk} =g {A_2}^{l,ijk} \Omega_{lm}\epsilon^m.
\end{align}
The gravitino variation gives a Killing spinor equation allowing for a solution with non-vanishing gravitino mass and thus for an anti-de Sitter vacuum.
From the second condition, we find 
\begin{equation}
\label{A2=0}
{A_2}^{l,ijk} = 0.
\end{equation}
As a consequence, the cosmological constant associated to a maximally supersymmetric anti-de Sitter vacuum is (we set $g=1$ from now on)
\begin{equation}
\label{VadsN=8}
V_{AdS} = -3 |A_1^{ij}|^2= -\frac25  \mathcal{M}^{MN}  {\rm Tr}\, ({\mathcal{Q}_M})^\dagger {\mathcal{Q}_N} ,
\end{equation}
where evaluation of the last expression at  ${A_2}^{i,jkl}=0$ is understood. Here, we used that the matrix ${\mathcal{Q}_{Mi}}^j$ is anti-hermitean and thus we introduced the notation ${{\mathcal{Q}_M}_i}^j {{\mathcal{Q}_N}_j}^i \equiv - {\rm Tr}\, ({\mathcal{Q}_M})^\dagger {\mathcal{Q}_N}$.

To determine the unbroken gauge group on the vacuum, we follow again \cite{Lust:2017aqj}. Given that there are no matter multiplets in the maximal theory, the group $H_{\rm mat}$ in the decomposition \eqref{GtoHRHmat} is vanishing and we are left with $H_R$. Again, this is identified as the group gauged by the graviphotons and leaving the gravitino mass matrix ${A_1}^{ij}$ invariant. For the maximal theory in five dimensions, $H_R = {\rm SU}(4)$ on a maximally supersymmetric anti-de Sitter vacuum, which indeed coincides with the R-symmetry group of the dual $\mathcal{N}=4$ SYM. (Similarly, for the maximal theory in four dimensions one has $H_R = {\rm SO}(4)$.) Therefore, with respect to the minimal theory, we have now an unbroken non-abelian gauge group on the vacuum. Notice that this group can be further reduced in case the vacuum preserves a lower amount of supercharges.

Having identified the gauge group, we proceed by repeating an analysis similar to that of section \ref{sec_noscalesepN=2} and thus generalizing \cite{Cribiori:2022trc} to five dimensions. First, we introduce the vielbeins $\Theta^m_M$, $\Theta_m^M$, with $\Theta^m_M \Theta_n^M = \delta^m_n$, such that
\begin{align}
\mathcal{M}_{MN} = g_{3/2}^{-2} \delta_{mn}\Theta^m_M \Theta^n_N,\qquad \mathcal{M}^{MN} = g_{3/2}^2 \delta^{mn} \Theta^M_m \Theta^n_N,
\end{align}
where the gravitino gauge coupling is
\begin{equation}
g_{3/2}^2 = \frac{1}{27}\, \delta_{mn} \Theta^m_M \mathcal{M}^{MN}\Theta^n_N.
\end{equation}
Then, we define the vectors (we omit the spacetime index in what follows)
\begin{equation}
\tilde A^m = \Theta^m_M A^{M}
\end{equation}
with canonically normalized kinetic terms
\begin{equation}
\begin{aligned}
-\frac{1}{16} \mathcal{M}_{MN} \mathcal{H}^M(A) \mathcal{H}^N(A) =-\frac{1}{16} g_{3/2}^{-2} \delta_{mn} \mathcal{H}^m(\tilde A) \mathcal{H}^n(\tilde A).
\end{aligned}
\end{equation}
We also define hermitian USp$(8)$ charges
\begin{equation}
{\tilde Q_{mj}}^{\,\,\,\,\,\,\,\,\,i}= -i\,\Theta^M_m {{\mathcal{Q}_M}_i}^j,
\end{equation}
such that the gravitino covariant derivative becomes
\begin{align}
D_\mu {\psi_\nu}^i =\dots + {A_\mu}^M {{\mathcal{Q}_M}_j}^i {\psi_\nu}^j= \dots +i {{\tilde A}_{\mu}}^{\,\,\,m} \,{\tilde Q_{mj}}^{\,\,\,\,\,\,\,\,\,i}  {\psi_\nu}^j.
\end{align}
Finally, we can rewrite the scalar potential \eqref{VadsN=8} as
\begin{equation}
\begin{aligned}
V_{AdS} = -\frac25 \mathcal{M}^{MN} {\rm Tr}\, {\mathcal{Q}_M}^\dagger \mathcal{Q}_N =-\frac25 g^2_{3/2} \,\delta^{mn}\, {\rm Tr} \,\tilde Q_m \tilde Q_n,
\end{aligned}
\end{equation}
or equivalently
\begin{equation}
\label{VadsN=8mod}
|V_{AdS}| = \frac25 g^2_{3/2} \,\delta^{mn}\, {\rm Tr} \,\tilde Q_m \tilde Q_n\, .
\end{equation}

As for the case of minimal supergravity, this shows in a model-independent way that the magnitude of the cosmological constant of maximally supersymmetric anti-de Sitter vacua is fixed by the gravitino gauge coupling.
Therefore, we can repeat the same arguments of sections \ref{sec_noscalesepN=2}, \ref{sec:holspecies} and conclude that scale separation is not possible in these vacua, for the Hubble radius and the Kaluza-Klein/species scale are of the same order.

The only difference with respect to the discussion in section \ref{sec_noscalesepN=2} is that now we are dealing with a non-abelian gauge group on the vacuum, $H_R = {\rm SU}(4)$.
We can nevertheless proceed by applying the weak gravity conjecture with respect to one of the abelian factors in the Cartan subalgebra of ${\rm SU}(4)$, having rank three. Concretely, without loss of generality, we can rotate the frame in such a way that the vector fields gauging the Cartan subalgebra are along $m=1,2,3$. For definiteness, we can choose then to apply the weak gravity conjecture with respect to, say, $\tilde A^1$ and thus from \eqref{VadsN=8mod} we can write
\begin{equation}
|V_{AdS}| \geq \frac25 \, g_{3/2}^2 \, {\rm Tr} \,\tilde Q_1 \tilde Q_1  \gtrsim \Lambda_{UV}^2.
\end{equation}
Then, the conclusions of section \ref{sec_noscalesepN=2} on the fate of scale separation follow.
Notice that one can apply the same logic to four-dimensional vacua. Hence, in this respect, the assumption of \cite{Cribiori:2022trc} on the need of an unbroken abelian gauge group on the vacuum can be relaxed also when dealing with the maximal theory.

\subsection{No effective description in de Sitter}

Finally, we present the extension of the argument of section \ref{sec:nodSN=2} to the maximal theory. Assuming that the gravitino mass is vanishing, ${A_1}^{ij}=0$, or at most parametrically small with respect to the Hubble scale, we have
\begin{equation}
\mathcal{M}^{MN}{{\mathcal{Q}_M}_i}^j {{\mathcal{Q}_N}_j}^i =-\frac23 |A_{2\,i,jkl}|^2
\end{equation}
and thus
\begin{equation}
V_{dS} = \frac13 |A_{2\,i,jkl}|^2   =\frac12 \mathcal{M}^{MN} \, {\rm Tr} {\mathcal{Q}_M}^\dagger \mathcal{Q}_N.
\end{equation}
Repeating the same steps as for the anti-de Sitter case, we can eventually write
\begin{equation}
\begin{aligned}
V_{dS} &= \frac12 g_{3/2}^2 \delta^{mn} {\rm Tr} \tilde Q_m \tilde Q_n\geq \frac12 g_{3/2}^2 \, {\rm Tr} \tilde Q_1 \tilde Q_1.
\end{aligned}
\end{equation}
This result is analogous to \eqref{VminimaldS2} and thus we can draw the same conclusions, namely de Sitter vacua with parametrically small gravitino mass are not trustworthy effective theories. In particular, upon enforcing the weak gravity conjecture, they lead to the problematic condition $\Lambda_{IR} \gtrsim \Lambda_{UV}$.

Searching for de Sitter critical points of the maximal theory is a challenging task even when one is not demanding for stability. To the best of our knowledge, no stable de Sitter vacuum is known at present. A systematic investigation of critical points of maximal supergravity with residual U$(2)$ symmetry in five dimensions has been recently carried out in \cite{Dallagata:2021lsc}. Among twelve critical points, only two are de Sitter with residual gauge group SO$(3) \times$SO$(3)$ and SU$(2)$ respectively. The first of these vacua was already found in \cite{Gunaydin:1985cu}, while the second is new. In \cite{Dallagata:2021lsc} the complete mass spectra are computed as well. We can thus see that, while the new vacuum features massive gravitini, the one already found in \cite{Gunaydin:1985cu} has vanishing gravitino masses and thus our analysis applies. In any case, as pointed out in \cite{Dallagata:2021lsc}, both these vacua are unstable.

\section{Conclusion}

In this work, we investigated from a bottom-up perspective two properties of effective field theories which are motivated from observations. These are the presence of a separation of scales between the size of the (unobserved) compact and the (observed) non-compact dimensions and the viability of an effective description of de Sitter vacua. We concentrated exclusively on five non-compact spacetime dimensions and we extended previous works performed in four dimensions \cite{Cribiori:2020use, DallAgata:2021nnr, Cribiori:2022trc}.

As for scale separation, we restricted our attention to supersymmetric anti-de Sitter vacua, for which we obtained model-independent results. We showed that scale separation is not possible in supersymmetric vacua of the minimal and the maximal theory as a consequence of some of the most established swampland conjectures, such as the weak gravity conjecture or the absence of global symmetries. We also provided an independent argument based on holography and on the species scale which points towards the same conclusion. We do not see any obvious obstruction in extending our analysis to vacua with an intermediate amount of preserved supercharges. Indeed, we believe that our results provide non-trivial evidence that supersymmetric anti-de Sitter vacua in five dimensions cannot be scale separated. In other words, their effective description, if anything, must be higher dimensional. In view of this fact, one could perhaps hope that non-supersymmetric vacua might still lead to genuinely five-dimensional effective theories with a separation of scale. Given that breaking supersymmetry hardly improves the behaviour of an effective theory, we believe this to be unlikely. Indeed, when combining our result with that of \cite{Ooguri:2016pdq}, stating that non-supersymmetric anti-de Sitter vacua are always unstable (this is motivated again from a sharpened version of the weak gravity conjecture), one can even speculate that any genuinely five-dimensional effective theory in anti-de Sitter belongs to the swampland, regardless of the number of preserved supercharges. 
We believe that this conclusion can be extended directly to higher dimensions, performing an analysis similar to that of the present work.

As for de Sitter vacua, we have been forced to introduce additional assumptions in order to present a model-independent argument. 
In particular, we needed to ask for the lagrangian gravitino mass to vanish on the vacuum.
While in four dimensions this condition is actually met in all known stable vacua of $\mathcal{N}=2$ supergravity, five-dimensional stable de Sitter solutions of minimal supergravity seem not to have it.
As a consequence, we could not rule them out in view of our argument and we hope to come back to this problem in the future. 
Let us just mention that, in view of the conjectures in \cite{Andriot:2022way,Andriot:2022xjh}, see also \cite{VanRiet:2011yc}, there seems to be little room to construct classical de Sitter vacua in five dimensions from a top-down perspective.

The analysis here presented can be extended in various directions. An immediate generalization for which we see no obvious obstruction is to look at supergravity in higher dimensions. Another interesting line of research would be to investigate further the holographic implications of our results concerning supersymmetric anti-de Sitter vacua. Furthermore, it would be important to better understand the connection between stable de Sitter vacua in four and five-dimensions, for the known ones seem to behave differently for what concern swampland conjectures. A first step was already taken in \cite{Ogetbil:2008tk}, but it is worth to pursue along this direction in view of its importance within the swampland program.

\paragraph{Acknowledgments.} We would like to thank G.~Dall'Agata, D.~L\"ust, E.~Palti and especially F.~Fa\-ra\-kos for discussions. 
The work of N.C.~is supported by the Alexander-von-Humboldt foundation.

\appendix
\section{On the scalar potential of maximal supergravity}
\label{app_N=8}

In this appendix, we report the necessary steps to recast the scalar potential of maximal supergravity,
\begin{equation}
V= \frac13 {A_{2\,i,jkl}}{{A_2}^{i,jkl}} - 3{A_{1\,ij}}{A_1}^{ij},
\end{equation}
in terms of the Killing vectors ${\mathcal{P}_M}^{ijkl}$ and of the gauge couplings ${\mathcal{Q}_{M\, i}}^j$, namely
\begin{equation}
V=\frac{1}{10} \mathcal{M}^{MN}{\mathcal{P}_{M\,ijkl}}{\mathcal{P}_{N}}^{ijkl} + \frac25 \mathcal{M}^{MN} {\mathcal{Q}_{M\, i}}^j {\mathcal{Q}_{N\, j}}^i.
\end{equation}
In practice, we derive the relations \eqref{Q2fin} and \eqref{P2fin}.

The main object to look at is the so-called $T$-tensor \cite{deWit:2004nw}, which intertwines ${\mathcal{P}_M}^{ijkl}$, ${\mathcal{Q}_{M\, i}}^j$ with ${A_1}^{ij}$, ${{A_2}^{i,jkl}}$  and it can be defined by appropriately dressing the embedding tensor with the scalar fields. For our purposes, it is sufficient to recall the two components of the $T$-tensor
\begin{align}
\label{TQV}
&{T^j}_{imn}  ={\mathcal{Q}_{M\,j}}^i {\mathcal{V}_{mn}}^M,\\
\label{TPV}
&{T^{ijkl}}_{mn} = {\mathcal{P}_M}^{ijkl} {\mathcal{V}_{mn}}^M,
\end{align}
which satisfy ${T^i}_{jkl}\Omega^{kl} = 0={T^{ijkl}}_{mn}\Omega^{mn} $ as a consequence of $\Omega^{mn}{\mathcal{V}_{mn}}^M=0$. The explicit definition of ${\mathcal{P}_M}^{ijkl}$, ${\mathcal{Q}_{M\, i}}^j$ is
\begin{align}
\label{QMconn}
{\mathcal{Q}_{M\, i}}^j &= \frac13 {\mathcal{V}_{ik}}^N \, {X_{MN}}^P {\mathcal{V}_{P}}^{jk},\\
\label{PMKilling}
{\mathcal{P}_M}^{ijkl} &= {\mathcal{V}_{mn}}^N {X_{MN}}^P {\mathcal{V}_{P}}^{[ij} \Omega^{k|m|} \Omega^{l]m}, 
\end{align}
where $X_M$ are the generators of the gauge group, namely $[X_M,X_N] = {f_{MN}}^P X_P$. The relations \eqref{TQV} and \eqref{TPV} can be inverted as
\begin{align}
\label{QVT}
&{\mathcal{Q}_{M\, j}}^i = {\mathcal{V}_{M}}^{mn} \,{T^i}_{jmn},\\
\label{PVT}
& {\mathcal{P}_M}^{ijkl} = {\mathcal{V}_{M}}^{mn} \,{T^{ijkl}}_{mn}.
\end{align}

Since the $T$-tensor is closely related to the embedding tensor, the constraints imposed on the latter force the former to be in the $\bm{351}$ representation of $E_{6(6)}$ as well. As a consequence of the decomposition $\bm{351}\to\bm{36}+\bm{315}$ under USp$(8)$, the components ${T^j}_{imn} $ and ${T^{ijkl}}_{mn}$ can be expressed in terms of ${A_1}^{ij}$, ${{A_2}^{i,jkl}}$ as
\begin{align}
\label{Tijkl}
{T_i}^{jkl} &= - \Omega_{im} {A_2}^{(m,j)kl} - \Omega_{im}\left(\Omega^{m[k}{A_1}^{l]j}+\Omega^{j[k}{A_1}^{l]m}+\frac14\Omega^{kl}{A_1}^{mj}\right),\\
\label{Tijklmn}
{T^{klmn}}_{ij} &= 4 {A_2}^{q,[klm}{\delta^{n]}}_{[i}\Omega_{j]q} + 3 {A_2}^{p,q[kl} \Omega^{mn]} \Omega_{p[i} \Omega_{j]q}.
\end{align}

We have now at our disposal all of the necessary ingredients to perform the computation. We start by deriving \eqref{Q2fin}. From \eqref{QVT} and using that ${T^i}_{jkl}$ is symplectic-traceless, we get
\begin{equation}
\label{MQQ=TT}
\mathcal{M}^{MN} {\mathcal{Q}_{M\, i}}^j {\mathcal{Q}_{N\, j}}^i = \Omega^{km}\Omega^{ln} \,{T^j}_{ikl} {T^i}_{jmn},
\end{equation}
where $\mathcal{M}^{MN}$ is the inverse of the matrix $\mathcal{M}_{MN}$ defined in \eqref{MMNkin}. Next, one can use \eqref{Tijkl} to express this in terms of ${A_1}^{ij}$ and ${{A_2}^{i,jkl}}$. There are four kinds of terms, which are schematically $A_1 A_1$, $A_1 A_2$, $A_2 A_1$ and $A_2 A_2$, but several simplifications occur due to the properties of ${A_1}^{ij}$, ${{A_2}^{i,jkl}}$, such as the fact that ${A_1}^{ij}={A_1}^{ji}$ or $A_{2\,i,jkl}=A_{2\, i,[jkl]}$, which imply in turn that $A_{2\,i,jkl}\,{A_1}^{kl}=0$. Eventually, the terms $A_1 A_2$ and $A_2 A_1$ (that are in fact one and the same) vanish identically because  ${{A_2}^{i,jkl}}$ is symplectic-traceless. Thus, one is left only with $A_1 A_1$ and $A_2 A_2$ terms, giving
\begin{equation}
\label{MQQ=TTfin2}
\mathcal{M}^{MN} {\mathcal{Q}_{M\, i}}^j {\mathcal{Q}_{N\, j}}^i =-\frac23{A_{2\,i,jkl}} {A_2}^{i,jkl}-\frac{15}{2} \,{A_{1\,ij}}{A_1}^{ij},
\end{equation}
which is precisely \eqref{Q2fin}. Next, we derive \eqref{PVT}. To this purpose, besides the aforementioned properties of  ${A_1}^{ij}$ and ${{A_2}^{i,jkl}}$, we need also the relation
\begin{equation}
\label{A22=A2A2}
{A_{2\,i,jkl}}\,{A_2}^{i,jkl} = 3 {A_{2\,j,ikl}}\,{A_2}^{i,jkl},
\end{equation}
which follows from $A_2^{[i,jkl]}=0$. From \eqref{PVT} and using that ${T^{klmn}}_{ij}$ is symplectic-traceless, we get 
\begin{equation}
\mathcal{M}^{MN}{\mathcal{P}_{M\,ijkl}}{\mathcal{P}_{M}}^{ijkl}  ={T^{ijkl}}_{mn}{T_{ijkl}}^{mn}.
\end{equation}
Next, one can use \eqref{Tijklmn} to express this in terms of ${{A_2}^{i,jkl}}$. Exploiting the properties mentioned before, together with \eqref{A22=A2A2}, after some steps one arrives at
\begin{equation}
\mathcal{M}^{MN}{\mathcal{P}_{M\,ijkl}}{\mathcal{P}_{M}}^{ijkl} = 6 A_{2\,i,jkl} {A_2}^{i,jkl},
\end{equation}
which is precisely \eqref{P2fin}.

\bibliography{references}  
\bibliographystyle{utphys}

\end{document}